\documentclass[10pt, conference, letterpaper]{IEEEtran}
\usepackage{algorithm}
\usepackage{algorithmicx}
\usepackage{algpseudocode}
\usepackage{amsfonts}
\usepackage{amsmath}
\usepackage{amssymb}
\usepackage{inputenc} 
\usepackage{xcolor}
\usepackage{mathtools}
\usepackage{graphicx}
\usepackage{caption}
\usepackage{subcaption}
\usepackage{import}
\usepackage{multirow}
\usepackage{cite}
\usepackage[export]{adjustbox}
\usepackage{breqn}
\usepackage{mathrsfs}
\usepackage{acronym}
\usepackage{setspace}
\usepackage{bm}
\usepackage{stackengine}
\usepackage{float}
\usepackage[font=small]{caption}
\usepackage{url}
\usepackage{tikz}

\renewcommand{\tablename}{Table} 

\usepackage{listings}

\lstdefinelanguage{BibTeX}
  {keywords={%
      @article,@book,@collectedbook,@conference,@electronic,@ieeetranbstctl,%
      @inbook,@incollectedbook,@incollection,@injournal,@inproceedings,%
      @manual,@mastersthesis,@misc,@patent,@periodical,@phdthesis,@preamble,%
      @proceedings,@standard,@string,@techreport,@unpublished%
      },
   comment=[l][\itshape]{@comment},
   sensitive=false,
  }

\usepackage{listings}

\renewcommand{\thetable}{\arabic{table}}

\graphicspath{{./figures/}}
\newcommand{\mytexttilde}{{\raise.17ex\hbox{$\scriptstyle\mathtt{\sim}$}}}

\title{Performance Evaluation of IoT LoRa Networks \\ on Mars Through ns-3 Simulations}

\author{\IEEEauthorblockN{{Manuele Favero, Alessandro Canova, Marco Giordani, Michele Zorzi}\medskip}
\IEEEauthorblockA{
Department of Information Engineering, University of Padova, Italy. \\
Email: \texttt{\{faveromanu,canovaales,giordani,zorzi\}@dei.unipd.it}\\
\vspace{-0.5cm}}}

\IEEEoverridecommandlockouts

\newcounter{remark}[section]

\begin{document}

\maketitle

\begin{abstract}
In recent years, there has been a significant surge of interest in Mars exploration, driven by the planet's potential for human settlement and its proximity to Earth. 
In this paper, we explore the performance of the LoRaWAN technology on Mars, to study whether commercial off-the-shelf IoT products, designed and developed on Earth, can be deployed on the Martian surface. 
We use the ns-3 simulator to model various environmental conditions, primarily focusing on the Free Space Path Loss (FSPL) and the impact of Martian dust storms. Simulation results are given with respect to Earth, as a function of the distance, packet size, offered traffic, and the impact of Mars' atmospheric perturbations. 
We show that LoRaWAN can be a viable communication solution on Mars, although  the performance is heavily affected by the extreme Martian environment over long distances.
\end{abstract}

\IEEEkeywords
LoRaWAN; Mars communication; Internet of Things (IoT); Network simulation; ns-3.
\endIEEEkeywords

\begin{tikzpicture}[remember picture,overlay]
		\node[anchor=north,yshift=-10pt] at (current page.north) {\parbox{\dimexpr\textwidth-\fboxsep-\fboxrule\relax}{
				\centering\footnotesize This paper has been accepted for presentation at the 2025 International Conference on Computing, Networking and Communications (ICNC). \textcopyright 2025 IEEE. \\
				Please cite it as: M. Favero, A. Canova, M. Giordani, M. Zorzi, ``Performance Evaluation of IoT LoRa Networks on Mars Through ns-3 Simulations,'' in International Conference on Computing, Networking and Communications (ICNC), 2025.\\
				}};
	\end{tikzpicture}

\section{Introduction}
\label{sec:introduction}

In recent years, Mars exploration has returned to the forefront of space research, with the Red Planet being widely considered as the most viable option for human colonization~\cite{NEUKART2024e26180}.
In fact, Mars is close to Earth, and has been extensively mapped and studied, particularly regarding its surface, water, and potential for life~\cite{marspast, wateronmars, carr2007surface}. 
Data collection efforts have been ongoing since the Mars Exploration Rover mission of NASA in 2004. 
More recent missions include:
NASA's Mars 2020, culminated with the rover Perseverance and the now-retired small robotic helicopter Ingenuity, landed in Jezero Crater in 2020 in search for traces of  water~\cite{perseverance}; the 2022 ExoMars mission, targeting sites likely to contain preserved organic material from Mars' early history~\cite{exomars}; China's Tianwen-1 and the UAE's EMM~\cite{uae} missions, aiming at further exploring and mapping the~planet.

\begin{figure}[t!]
\begin{minipage}{0.34\textwidth}
\centering
\includegraphics[height=0.8\textwidth]{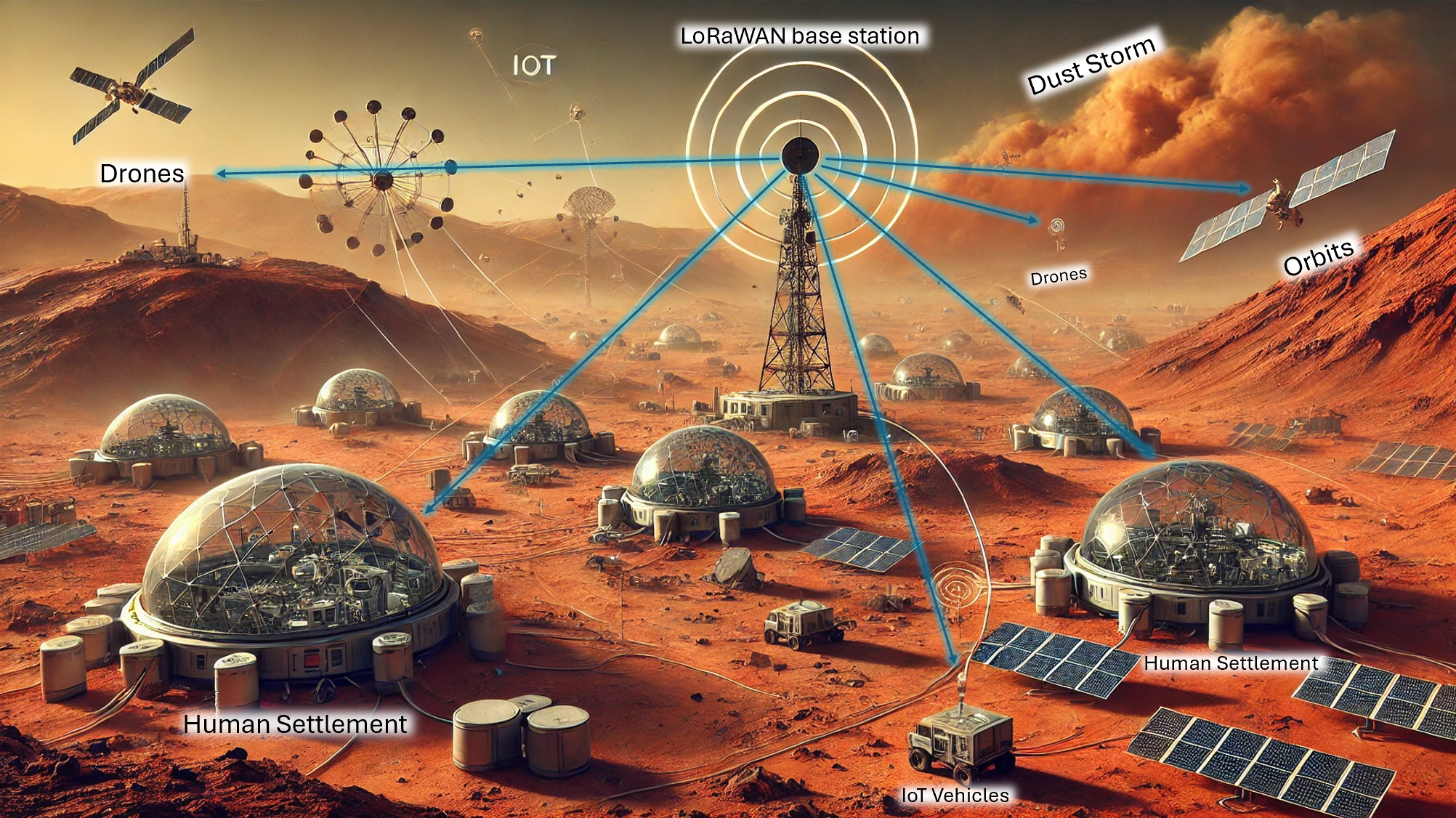}
\end{minipage} \hspace{1.8cm}
\hspace{0.1cm}
\caption{An illustration of a futuristic telecommunication scenario on Mars with IoT devices communicating through LoRa.}
\label{fig:marsia}
\end{figure}

From a telecommunications perspective, these missions have provided precise data for modeling the Martian communication channels, including both air-to-ground and ground-to-ground links~\cite{10115642, bell2000mars}. 
Future goals also include sending humans to Mars and establishing long-term settlements, as shown in Fig.~\ref{fig:marsia}. 
Notably, future networks on Mars are likely to integrate ground nodes and infrastructure with drones, CubeSats, and orbiters to provide on-planet connectivity and connectivity to Earth. For instance, Bonafini \emph{et al.} presented a 3D network architecture that combines  drones, CubeSats, and the Cloud-Radio Access Network (C-RAN), highlighting the critical role of orbital design to optimize communication performance on Mars~\cite{bonafini2020building, bonafini2022end}. 

Achieving this ambitious vision requires the design of robust telecommunication systems. 
While Long Term Evolution (LTE) networks have been already explored for Martian communication, showing potential for short-range links, further enhancements are required to address the high path loss experienced over long distances on Mars~\cite{bonafini2020building}.
In this context, the Internet of Things (IoT) represents a promising solution for Mars communication~\cite{iotsc}. IoT technologies, especially Long Range (LoRa), offer several key advantages tailored to the Martian environment, including: (i) long-range communication to cover vast distances; (ii) low energy consumption for battery-powered devices without a direct power supply; (iii) robustness against interference and noise; and (iv) scalability to support a large number of nodes per gateway.
In confirmation of this, IoT applications based on LoRa have already emerged for Mars. For example, Ledesma \emph{et al.} suggested a satellite-IoT architecture using LoRa to connect sensors and orbiters \cite{ledesma2024architectural}. 
On Earth, LoRa (with LoRaWAN) networks have shown great scalability~\cite{haxhibeqiri2017lora, to2018simulation, chen2023multi}, which motivates research toward the study of this technology to address the unique communication challenges of Mars.

Along these lines, in this paper we evaluate the feasibility of deploying an IoT system based on the LoRaWAN protocol stack on Mars.
The goal is to identify performance gaps with respect to Earth, and provide guidelines for adapting commercial off-the-shelf IoT products to work on Mars accordingly.
To do so, we extend the LoRaWAN simulation module for ns-3, developed and maintained by the University of Padova, by implementing a Martian channel model based on real data.
Specifically, we incorporate the effects of the atmosphere on Mars, and the presence of dust storms and particles.
Simulation results, as a function of the communication distance, packet size,
offered traffic, and the impact of Mars' atmospheric perturbations, indicate that LoRa is feasible for Martian IoT applications, although the performance is severely deteriorated with respect to Earth.
In particular, the normalized throughput is similar at short distances, but decreases by approximately 50\% at longer distances. Specifically, we obtain the same throughput on Mars at around one-third of the distance than on Earth.

The rest of the paper is organized as follows.
In Sec.~II we describe the Martian channel model. In Sec. III we describe the features of LoRa and our ns-3 module implementation. In Sec. IV we present our simulation results. In Sec. V we conclude the paper with suggestions for future work.

\section{Description of the Martian Channel}
\label{sec:Model_Description}
Telecommunications are crucial for the success of in-planet missions. Notably, it is essential to study how the Martian environment affects radio wave propagation, especially in comparison to Earth.
In this section we present the features of the Mars atmosphere (Sec.~\ref{sub:athm}), and describe a simple and tractable channel model for communication based on data from Mars missions (Sec.~\ref{sub:channel}).

\subsection{Mars Atmosphere}
\label{sub:athm}
The main aspect that affects wireless communications on Mars is its unique atmosphere. 
Thanks to the data collected by rovers and robots on the planet during the past years, we know that the atmosphere of Mars is very different from Earth's. The main difference is that the Martian atmosphere extends for approximately 100 Km, compared to around  10\,000 km  on Earth.
Consequently, the atmospheric pressure on Mars is extremely low, i.e., around 6.1 mbar, which is only 0.6\% of the atmospheric pressure on our planet. A thin atmosphere influences the air temperature.
Indeed, unlike Earth, Mars' atmosphere cannot effectively retain heat from solar radiation: as a result, temperatures vary widely, from 140 K at the poles in winter to 296 K at the equator in summer~\cite{barth1974atmosphere}.

Most importantly, Martian winds can frequently cause large-scale dust storms, particularly during late spring or early summer in the southern hemisphere when Mars is at perihelion.
These storms can vary in intensity, from mild to severe, and can drastically impact the features of the Martian surface. Specifically, they pose challenges for wireless communications by potentially disrupting radio wave propagation, especially due to the scattering and absorption of signals by fine, electrically charged dust particles~\cite{ho2002radio}. These particles create a dynamic electromagnetic environment that can distort signals, cause fading, and lead to multi-path propagation. Charged particles and atmospheric turbulence during storms can also interfere with the signal, while dust accumulation on antennas may degrade the communication performance.

\subsection{Channel Model}
\label{sub:channel}
In general, wireless signal propagation depends on several factors~\cite{goldsmith2005}, especially path loss. In free space, path loss only depends on the carrier frequency and the distance between the transmitter and the receiver. In more complex environments, like in urban or indoor scenarios, it further increases due to physical obstacles like buildings or terrain. 
Wireless signals are also affected by fading, which refers to the time-varying fluctuations of the signal amplitude due to multi-path propagation, delay spread, and interference.

As described in Sec. II-A, the main components that affect wireless communications on Mars are its atmosphere and dust storms. Indeed, in this study we decided to ignore secondary factors such as fading, interference, delay spread and multi-path propagation. Rather, we model the channel in terms of the Free-Space Path Loss (FSPL) $L(d)$, as a function of the distance $d$ between the transmitter and the receiver, and the attenuation $A_{ds}(\lambda)$ due to dust storms, as a function of the wavelength $\lambda$. Specifically, the total path loss on Mars can be written as 
\begin{equation}
\label{eq:pathlossMars}
    PL\,[dB] = 10\log(L(d))+A_{ds}(\lambda) \cdot d.
\end{equation}

\paragraph{Free-space path loss}
For the Martian terrain, characterized by a medium-high density of rocks and numerous scatterers, the FSPL is modeled using a third-order exponent as described in~\cite{perf_evaluation}, compared to Earth where we have a second-order exponent. Therefore
\begin{equation}
\label{eq:spreadingFactor}
    L(d) = \Big({4\pi d}/{\lambda} \Big)^3,
\end{equation}
with $\lambda={c}/{f}$, where $c$ is the speed of light and $f$ is the carrier frequency.

\paragraph{Attenuation due to dust storms}
Dust storms are a common weather phenomenon on the Martian surface, particularly in the southern hemisphere, though they can extend across the entire planet. 
These storms are driven by strong winds that lift dust particles into the atmosphere~\cite{perf_evaluation}, and the resulting impact on wireless communications can be significant, especially for high-frequency signals.

To assess the effects of dust storms, in this paper we consider three levels of intensity, defined by the size of the dust particles.
Specifically, the attenuation due to dust storms is modeled as described in \cite{bonafini2020building}, i.e.,
\begin{equation}
\label{eq:duststorm}
    A_{ds}(\lambda) = \frac{
        1.029 \times 10^6 \epsilon^{''}
    }{
        \lambda \cdot [(\epsilon^'+2)^2 + \epsilon^{''2}]
    }N_T\Bar{r}^3.
\end{equation}

Notably, $A_{ds}$ depends on the wavelength $\lambda$, the real and imaginary parts of the dielectric permittivity of dust particles $\epsilon^'$ and $\epsilon^{''}$, respectively, the total density of particles per unit volume $N_T$, and the mean particle radius $\bar{r}$. 
The numerical values of these parameters will be reported in Sec. IV.





\section{Description of LoRa and Ns-3 Simulations}
In this section we review the main features of LoRa for IoT (Sec.~\ref{sub:lora}), and present our ns-3 module for the simulation of LoRa networks on Mars (Sec.~\ref{sub:ns-3}).

\subsection{LoRa and LoRaWAN}
\label{sub:lora}
LoRa is a proprietary Low-Power Wide Area Network (LPWAN) technology developed by Semtech. It operates in the lower ISM bands (EU: 863-870 MHz, USA: 902-928 MHz), with a bandwidth of 125 kHz and a maximum transmit power of 14 dBm.
It is based on Chirp Spread Spectrum (CSS) modulation, where symbols are encoded by modulating a carrier that changes frequency linearly in time.
LoRa uses the Spreading Factor (SF) as a key parameter, ranging from 7 to 12,  where $2^{SF}$ is the number of channel symbols used to transmit each information bit. The SF determines the trade-off between data rate and communication range: as the SF increases, transmissions take more time, which reduces the sensitivity requirements at the receiver (from $-132$ dBm with SF7 to $-143$ dBm with SF12) and increases the coverage (up to 14 km with SF12), at the expense of a lower data rate (from 5.5 Kbps with SF7 to 0.29 Kbps with SF12)~\cite{LORA2}.
Generally, the SF is assigned based on the power level, where each device uses the lowest possible SF such that the received power is above the sensitivity of the receiver. 
The combination of CSS with SF guarantees robustness to interference. In fact, multiple signals at different data rates (i.e., using different SFs) on the same channel (i.e., using the same time/frequency resources) can arrive concurrently with no collision.

Formally, LoRa operates at the physical layer, and is usually combined with {LoRaWAN} for the implementation of the rest of the protocol stack, especially the Medium Access Control (MAC) layer.
LoRa networks generally define a star topology, where nodes communicate directly with a central entity, i.e., the gateway, connected to a fixed network infrastructure. This architecture is suitable for applications like data collection~\cite{LORA1}, and is particularly relevant to scenarios such as those on Mars, especially to connect sensors (e.g., on robots or rovers) with distant receivers (e.g., drones, orbiters, or gateways on the base campus).

\subsection{An Ns-3 Module for LoRa Networks on Mars}
\label{sub:ns-3}

Ns-3 is a discrete-event network simulator widely used for research and education in networking. 
While most simulators focus on Physical (PHY) and Medium Access Control (MAC) layer designs, and sacrifice the accuracy of the higher layers to reduce the computational complexity, ns-3 incorporates accurate models of the whole protocol stack, thus enabling scalable and realistic end-to-end simulations. 
For this work we used the \texttt{lorawan} ns-3 module~\cite{7996384}, developed as an open-source framework by the University of Padova.\footnote{The source code of the \texttt{lorawan} module is available here: \url{https://github.com/signetlabdei/lorawan}.} 
This module implements a set of classes for modeling the modulation and medium access technology of a LoRaWAN network.
Thanks to a simple underlying physical layer model, 
this module can support simulations with a large number of devices that access the wireless channel infrequently, like in IoT scenarios. 
Within this module, we implemented the \texttt{MarsPropagationLossModel} class by directly modifying the \texttt{PropagationLossModel} class in the ns-3 source, adding code to describe the Martian channel as detailed in Sec.~\ref{sub:channel}.\footnote{The source code of the Mars channel model simulator is publicly available here: \url{https://github.com/manuelefavero/MarsPropagationLoss}.} We then modified the \texttt{AlohaThroughput} class in the LoRaWAN module by integrating Martian-specific variables, and making them externally accessible and adjustable. Simulation are run using the Python \texttt{sem} library,\footnote{The source node of the \texttt{sem} library is available here: \url{https://github.com/signetlabdei/sem}.} which permits to execute multiple ns-3 scripts, and manage and collect results in processing-friendly data structures.

\section{Performance Evaluation}
\label{sec:ns3sim}

\begin{table}[t!]
\renewcommand{\arraystretch}{1.3}
\footnotesize
\centering
\caption{Simulation parameters.}
\label{tab:params}
\begin{tabular}{|l|l|}
\hline
\textbf{Parameter}                              & \textbf{Values}                 \\ \hline
$\Re$ dielectric permittivity of dust particles ($\epsilon^'$) & 2.9038 C$^2$N$^{-1}$m$^{-2}$ \\
$\Im$ dielectric permittivity of dust particles ($\epsilon^{''}$) & 
0.1278 C$^2$N$^{-1}$m$^{-2}$ \\\hline
Total density of particles per volume unit ($N_T$) & \begin{tabular}[c]{@{}l@{}}$1\cdot 10^6$ m$^{-3}$\\ $1\cdot 10^7$ m$^{-3}$\\ $3\cdot 10^7$ m$^{-3}$\end{tabular} \\\hline
Mean particle radius ($\bar{r}$) & \begin{tabular}[c]{@{}l@{}}$1.5\cdot 10^{-6}$ m$^{-3}$\\ $4.5\cdot 10^{-6}$ m$^{-3}$\\ $20\cdot 10^{-6}$ m$^{-3}$\end{tabular} \\\hline
Frequency $(f)$ & $868$ MHz
\\\hline
Bandwidth  $(B)$ & $125$ KHz
\\\hline
Number of end nodes  $(n)$ & $1000$ 
\\\hline
Transmission Power  $(P_{\rm tx})$ & $14$ dBm
\\\hline
Min. LoRa throughput ($S_m$) \cite{adelantado2017understanding} & $300$ bit/s \\
\hline
\end{tabular}
\vspace{-0.2cm}
\end{table}


In this section we present our ns-3 results using the simulation module described in Sec.~\ref{sub:ns-3},  to evaluate the performance of LoRa on Mars, in comparison with Earth. The list of simulation parameters is available in Table~\ref{tab:params}.

\begin{figure*}[t!]
\centering
\begin{minipage}{0.4\textwidth}
  \centering
  \includegraphics[height=0.8\textwidth]{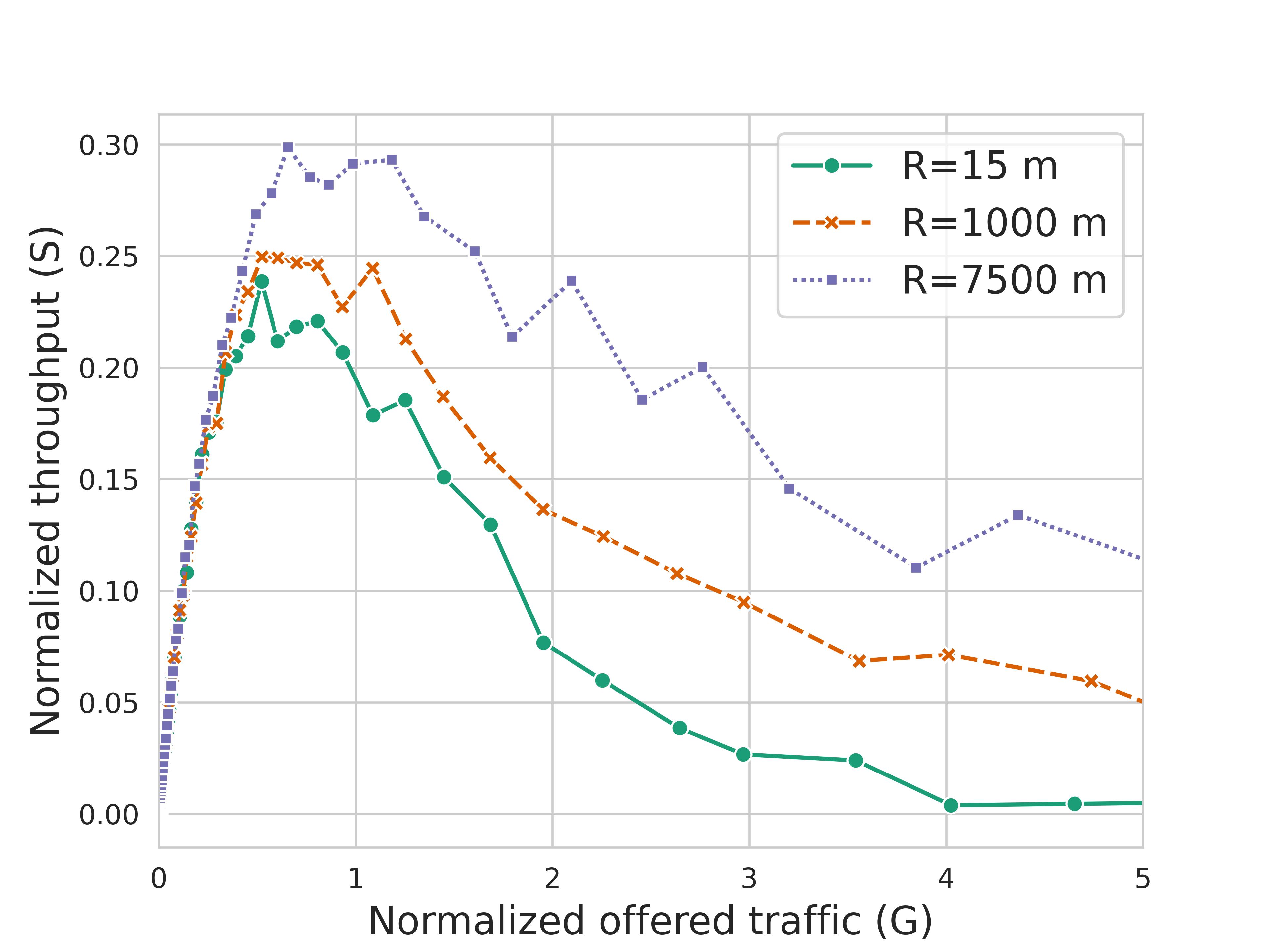}
  \subcaption{Earth scenario.}
  \label{fig:earth}
\end{minipage} \hspace{1cm}
\begin{minipage}{0.4\textwidth}
  \centering
  \includegraphics[height=0.8\textwidth]{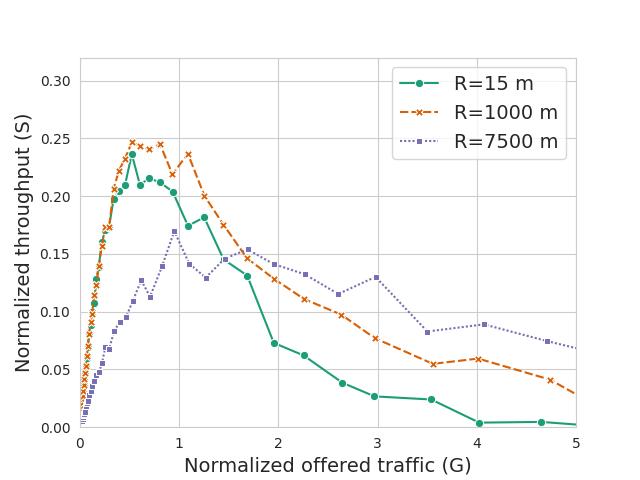}
  \subcaption{Mars scenario.}
  \label{fig:mars}
\end{minipage}

\caption{Normalized throughput ($S$) on Mars and on Earth vs. the normalized offered traffic ($G$) and $R$, in a scenario with $n=1000$ IoT end nodes sending packets of size 50 Bytes.\vspace{-0.33cm}}
\label{fig:comparison2}
\end{figure*}

\begin{figure*}[t!]
\centering
\begin{minipage}{0.6\textwidth}
\centering
\includegraphics[height=0.9\textwidth]{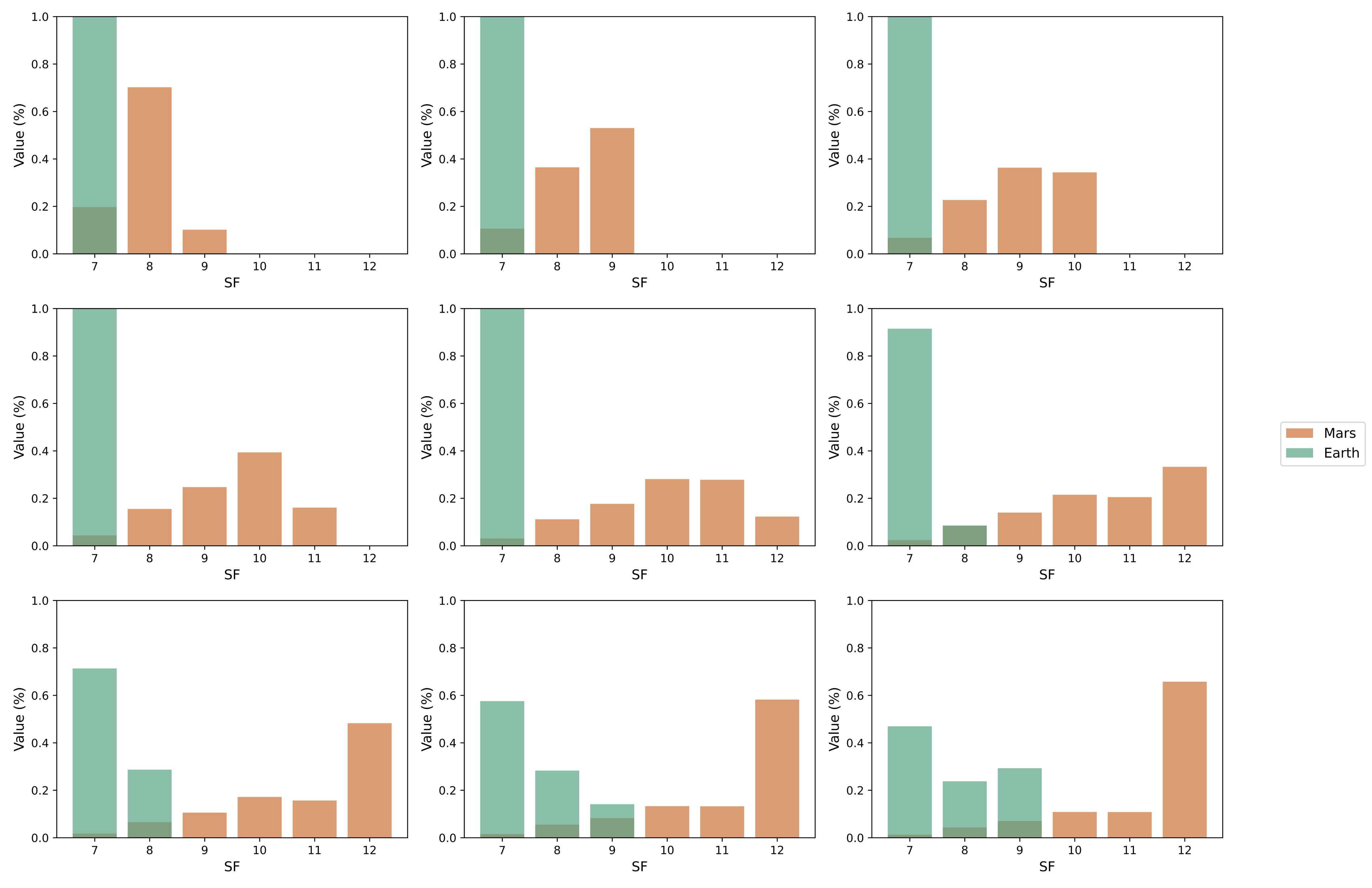}
\end{minipage} \hspace{3.5cm}
\hspace{0.1cm}
\caption{SF distribution on Earth vs. Mars for different values of the radius of the deployment area, from $R=1000$ m (top-left corner) to $R=4000$ m (bottom-right corner), and for $d=1000$ m.\vspace{-0.33cm}}
\label{fig:sf_variation}
\end{figure*}

The scenario consists of a LoRaWAN gateway at a distance $d$ from the center of a circle of radius $R$, where $n$ IoT nodes (e.g., sensors, robots, rovers, etc.) are deployed according to a Poisson Point Process (PPP) in a fixed position, so they do not move during the simulation.
End nodes communicate with the gateway using Slotted-ALOHA as transmission protocol, sending packets with a rate that depends on the packet size and the inter-arrival time.
Based on the considerations in~\cite{adelantado2017understanding}, we set the minimum required throughput for LoRa end nodes to $S_m=300$ bit/s.

As far as the dust storm attenuation model presented in Eq.~\eqref{eq:duststorm} is concerned, the values of $\epsilon^'$ and $\epsilon^{''}$ have been estimated by interpolation considering data measurements from real rovers sent to Mars. We consider three different types of dust, namely CalicheA, CalicheB and Sand~\cite{dustEpsilonEstimation2,dustEpsilonEstimation1}.
In turn, $N_T$ and $\Bar{r}$ depend on the intensity of the dust storms, and we consider low, moderate, and severe intensity. 
In general, increasing the intensity of the storm affects the average radius of the dust particles $\bar{r}$ within the storm. This occurs because the intensity of a dust storm depends on the strength of the winds that generate it: the stronger the wind, the larger the particles lifted into the atmosphere, which generates more severe attenuation.

\begin{figure}[t]
\centering
\includegraphics[width=0.4\textwidth]{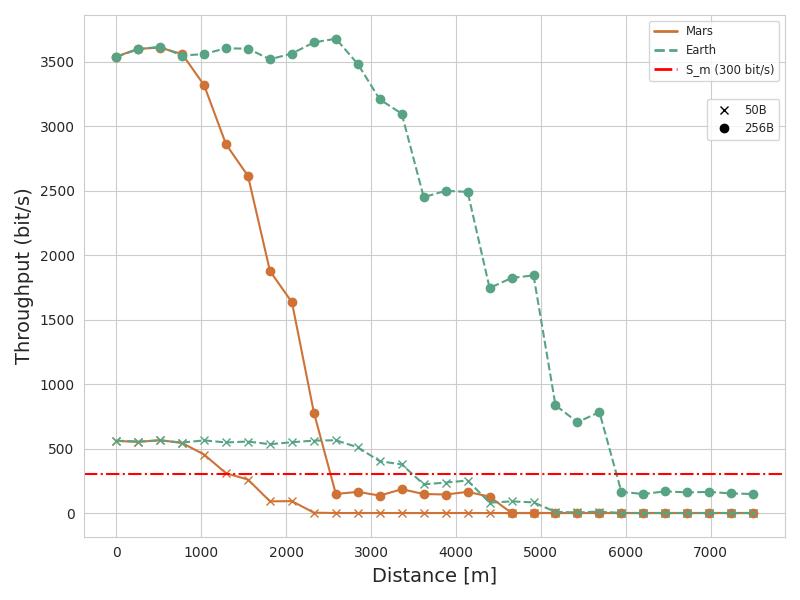}
\label{fig:raten}
\hspace{0.1cm}
\caption{Throughput on Mars and on Earth vs. $d$ and the packet size.\vspace{-0.33cm}}
\label{fig:packetcomparison}
\end{figure}

\subsection{Results: Earth vs. Mars} 
First, we compare the performance of an IoT LoRa network with $n=1000$ end nodes on Earth vs. Mars. Using Slotted-ALOHA, we analyze the relationship between the normalized throughput ($S$) and the normalized offered traffic ($G$, a dimensionless metric defined as the normalized ratio of the number of transmission attempts and the channel capacity), as shown in \mbox{Fig.~\ref{fig:comparison2}}. 
On Earth, $S$ reaches a maximum value of 0.3, close to the theoretical maximum of 0.37 for Slotted-ALOHA~\cite{roberts1975aloha}, with the best performance observed for a deployment radius of $R=7500$ m.
In fact, as $R$ increases,  the $n$ end nodes are sparser, and SFs are more uniformly distributed, which reduces the collision probability.
On Mars, instead, $S$ is limited to 0.26, and the maximum range is also limited with respect to Earth to only $R=1000$ m.
These results indicate that the impact of the FSPL on the Martian channel is significant, and imposes relative small-scale IoT deployments. 
In \mbox{Fig.~\ref{fig:sf_variation}} we illustrate the SF distribution for end nodes on Earth and on Mars. On Earth, propagation conditions are good, and devices primarily use SF7 to maximize the data rate, while SF8 and SF9 are selected only when $R>3000$ m. 
Conversely, on Mars, end nodes often use SF12 to operate at much lower sensitivity levels and compensate for the additional path loss. While this approach permits to reduce the probability of collision since more SFs are selected, it also decreases the data rate, and so the throughput.

In \mbox{Fig.~\ref{fig:packetcomparison}} we compare the average throughput on Earth and Mars as a function of the packet size and the distance $d$ from the LoRa gateway, for $R=1000$ m and a simulation time of 500 seconds. 
On Earth, with packets of 50 Bytes (i.e., $G=0.8$ Kbit/s), the 
max. communication distance at which end nodes can be deployed to satisfy the minimum required throughput $S_m$ is $\bar{d}=3500$ m. On Mars, under the same conditions, this distance is limited to only~$\bar{d}=1500$ m. Increasing the packet size to 256 Bytes (i.e., $G=4.096$ Kbit/s), the throughput improves in both scenarios. However, on Mars, we have $\bar{d}=2500$ m, whereas on Earth it is $\bar{d}=6000$ m.  

In general, we can evaluate the performance gap of LoRa on Earth vs. Mars. Specifically, the maximum normalized throughput on Mars is around 15\% lower than on Earth. Notably, we obtain the same throughput on Mars at around one-third of the distance than on Earth, which may complicate network deployment.


\subsection{Results: Different Martian Scenarios} 

 \begin{figure}[t!] 
    \centering
    \includegraphics[width=0.4\textwidth]{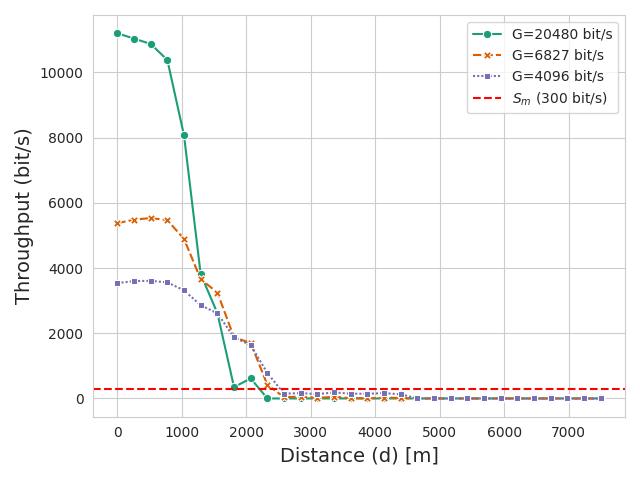}
    \caption{Throughput on Mars vs. $d$ and $G$, with a packet size of 256~B.\vspace{-0.1cm}}  
    \label{fig:svsrates} 
\end{figure}

\begin{figure}[t!]
\centering
\includegraphics[width=0.5\textwidth]{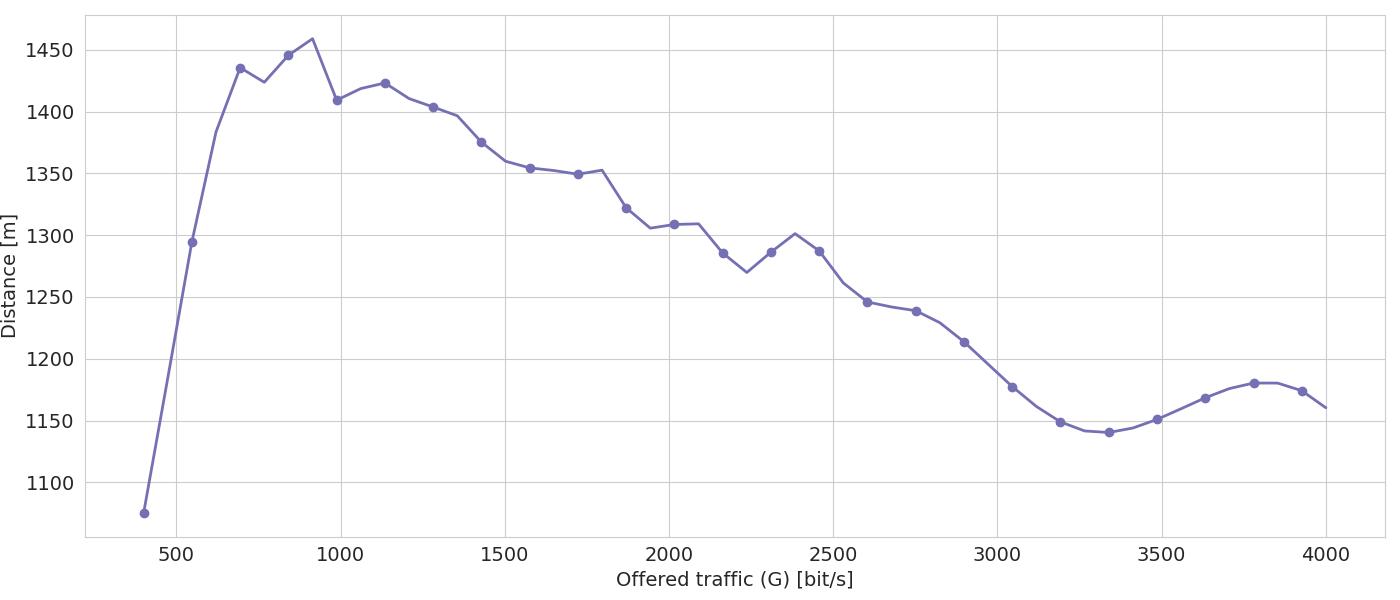}
\label{fig:raten}
\hspace{0.1cm}
\caption{Max. distance $\bar{d}$ on Mars to achieve the minimum throughput $S_m$ vs. $G$, with a packet size of 50 B.}
\label{fig:distvstraffic}
\end{figure}

Next, we focus on Mars simulations to evaluate the impact of the offered traffic $G$ and the distance from the gateway $d$ on the achievable throughput of a LoRa network, with a packet size of 256 Bytes.
As expected, in \mbox{Fig. \ref{fig:svsrates}} we observe that the throughput initially increases as $G$ increases. However, this comes with some drawbacks. 
In fact, for $G=20.48$ Kbit/s, despite the higher initial throughput, the performance rapidly drops as $d>1200$ m given the severe impact of atmospheric attenuation, and $\bar{d}=1800$ m. This behavior is likely due to a higher probability of collision as $G$ increases. 
Conversely, for $G=6.827$ Kbit/s we have more stable performance, and $\bar{d}$ is up to 2500 m. However, the maximum throughput in this case is only 6 Kbps, i.e., around 50\% lower than for $G=20.48$ Kbit/s.
These results highlight a trade off: the throughput is maximized at short distances, but deteriorates rapidly as the distance between the LoRa gateway and the end nodes increases. For $d>1500$ m, it is necessary to reduce the offered traffic, e.g., by decreasing the packet size and/or the inter-arrival time, to maintain stable performance.

Furthermore, in \mbox{Fig. \ref{fig:distvstraffic}} we plot $\bar{d}$ as a function of $G$, and considering a packet size of 50 Bytes.
At first, when $G$ is small, collisions are rare, so $\bar{d}$ increases, peaking at approximately 1450 m for $G=1000$ bit/s. However, as $G$ continues to increase, the probability of collision also increases, so $\bar{d}$ starts to decrease.
This behavior confirms our previous observations, and aligns with the limitations of Slotted-ALOHA used for transmissions in this scenario.

 \begin{figure}[t] 
    \centering
    \includegraphics[width=0.4\textwidth]{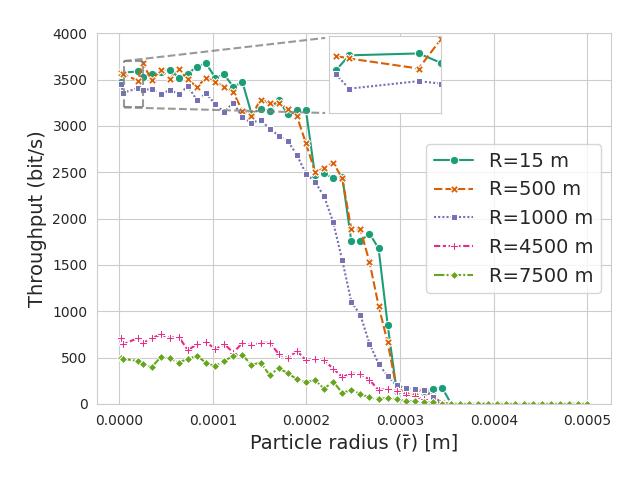}
\vspace{-0.33cm}    \caption{Throughput on Mars vs. $R$, as a function of the size $\bar{r}$ of dust particles. The zoom refers to the typical values of $\bar{r}$ on~Mars.}  
    \label{fig:dust} 
\end{figure}

\subsection{Results: Impact of Dust Storms} 

Finally, in \mbox{Fig. \ref{fig:dust}} we analyze the impact of dust storms on the Martian surface for IoT networks. 
Dust storms are particularly characterized by the radius of sand particles $\bar{r}$, as described in \mbox{\eqref{eq:duststorm}}. 
 In general, we see that the effect of $\bar{r}$ is significant. However, for the typical values of $\bar{r}$ in the Martian environment as reported in Table~\ref{tab:params}, where $\bar{r}$ is up to $20\cdot10^{-6}$ m for severe dust storms, the impact is negligible (see the zoom in the top-left corner of Fig. \ref{fig:dust}), and can be disregarded.
This is because LoRa operates at a relatively low frequency, i.e., 868 MHz in our simulations, where signal propagation is not affected by small particles. 

\begin{figure}[t] 
    \centering
    \includegraphics[width=0.4\textwidth]{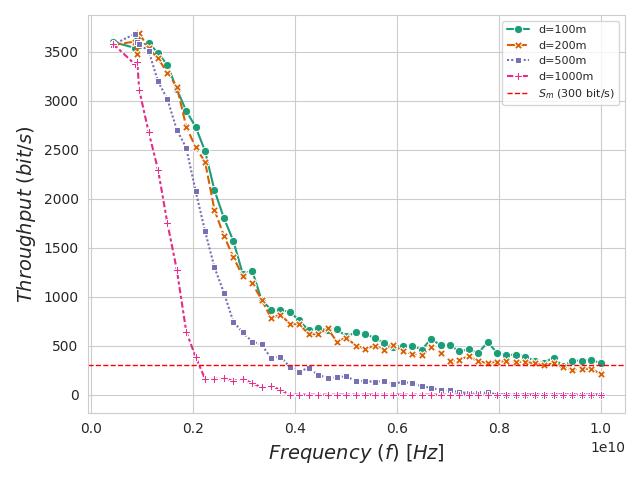}
    \caption{Throughput on Mars vs. $f$ in a severe dust storm scenario (i.e., $N_T= 3\cdot 10^7$ m$^{-3}$, $\bar{r}=20\cdot10^{-6}$ m), as a function of $d$.}  
    \label{fig:dustfrequencies} 
\end{figure}

For completeness, we illustrate in Fig. ~\ref{fig:dustfrequencies} the impact of particles as a function of the operating frequency $f$. We observe that $f$ has a strong effect on the throughput: while for LoRa the impact is negligible (the throughput is around 3 Kbit/s at 868M Hz), communication in the 2.3 GHz spectrum, where other telecommunication services such as Wi-Fi and LTE typically operate, may be problematic (the throughput drops to less than 1 Kbit/s, especially at long distances). 
We conclude that the effect of dust storms is limited for LoRa networks. Still, dust particles may obscure antennas and drastically reduce the power of solar panels, thus creating additional issues that should be properly addressed~\cite{he2024martian}.

\section{Conclusions}
\label{sec:conclusions}
In this study we used ns-3 to simulate an IoT communication system on the Martian surface based on the LoRaWAN technology. The goal was to assess whether such a system could be deployed on Mars, and to evaluate its performance under the same configurations used on Earth.
We modeled the Martian channel in terms of the FSPL and attenuation caused by dust storms, incorporating these factors into ns-3.
Our results showed that the throughput on Mars decreases faster with the distance compared to Earth. However, relatively long-range communication, i.e., up to around 2000 m, can still be achieved by reducing the offered traffic. Additionally, we observed that, at the frequencies used by LoRa, dust storms do not significantly affect the system performance. 

Further research is needed to improve the channel model by incorporating additional effects, such as the impact of delays caused by multi-path effects.

\section*{Acknowledgment}
This research was partially supported by the European Union under the Italian National Recovery and Resilience Plan (NRRP) of NextGenerationEU, 
partnership on ``Telecommunications of the Future'' (PE0000001 - program “RESTART”).



\bibliography{biblio}
\bibliographystyle{ieeetr}

\end{document}